\documentclass{iau} 
\usepackage{graphicx}

\title[Multiple populations in NGC~3201] 
{Breaking the dichotomy between typical and anomalous globular clusters: the case of NGC~3201}

\author[Bruno Dias et al.]   
{Bruno Dias$^{1,2}$,
Ignacio Araya$^{3}$,
Jo\~ao Paulo Nogueira-Cavalcante$^{4}$,
Leila Saker$^{5}$,
 \and Ahmed Shokry$^{6}$}

\affiliation{$^{1}$European Southern Observatory, Alonso de C\'ordova 3107, Vitacura 19001, Chile \\ [\affilskip]
$^{2}$Departamento de F\'{\i}sica, Facultad de Ciencias Exactas, Universidad Andr\'es Bello, Av. Fernandez Concha 700, Las Condes, Santiago, Chile\\ [\affilskip]
$^{3}$Núcleo Matem\'aticas, F\'isica y Estad\'istica, Facultad de Estudios Interdisciplinarios, Universidad Mayor, Manuel Montt 318, Providencia, Santiago, Chile\\ [\affilskip]
$^{4}$Observat\'orio Nacional, Rua General Jos\'e Cristino 77, Rio de Janeiro, RJ 20921-400, Brazil \\ [\affilskip]
$^{5}$Observatorio Astron\'omico de C\'ordoba, Laprida 854, 5000, C\'ordoba, Argentina\\ [\affilskip]
$^{6}$National Research Institute of Astronomy and Geophysics (NRIAG), 11421, Helwan, Cairo, Egypt
}

\pubyear{2019}
\volume{351}  
\jname{Star Clusters: From the Milky Way to the Early Universe}
\editors{A. Bragaglia, M.B. Davies, A. Sills \& E. Vesperini, eds.}
\begin{document}

\maketitle

\begin{abstract}
We recently discovered that NGC~3201 has characteristics
  that set it outside the current twofold classification scheme for
  Galactic globular clusters (GCs).  Most GCs are mono-metallic and show
  light-element abundance variations (e.g., Na-O and C-N
  anti-correlations); but a minority of clusters also
  present variations in Fe correlating with s-process element and C+N+O
  abundances, and they possess {\em multiple} C-N sequences. These
  anomalous GCs also have a broad sub-giant branch (SGB) and follow the
  same mass-size relation as dwarf galaxies possibly
  evolving into GCs.  We now revealed that NGC~3201 belongs to neither
  group.  It has multiple C-N sequences, but no broad
  SGB, no strong evidence of a Fe-spread, and it does not follow the mass-size relation. 

\end{abstract}

\firstsection 

\section{Introduction}

Typical globular clusters (GCs) are mono-metallic and present star-to-star variation of light-element abundances as detected by high-/low-resolution spectroscopy (e.g. Na-O, C-N anti-correlations, Briley et al. 1994, Gratton et al. 2004) and high-precision UV photometry (e.g. chromosome maps, Milone et al. 2017). A minority of peculiar clusters also present abundance variations in Fe, s-process elements, C+N+O, correlating with multiple C-N sequences; they also show a broad sub-giant branch (SGB) and follow a common mass-size relation (e.g. Marino et al. 2015, Da Costa 2016, Dias et al. 2018). The nonstop saga to explain how some polluters enrich second generation stars of typical GCs in light elements but not in Fe has achieved substantial contributions but still no model can fully explain all observational evidences (e.g. Bastian \& Lardo 2018). For the peculiar clusters the puzzle is more complex and some have tried to explain the origin of the peculiar GCs possibly as the nucleus of captured dwarf galaxies (e.g. Bekki \& Freeman 2003). 

Recently, two GCs challenged this dichotomy. NGC~6934 was analysed by Marino et al. (2018) using high-resolution spectroscopy and they found a Fe-spread, in agreement with the chromosome map, but no variation in s-process elements, therefore sharing characteristics of typical and anomalous GCs. NGC~3201 was analysed by us using low-resolution spectroscopy (Dias et al. 2018). We found three sequences of CN-CH anti-correlation (or a broad correlation in the interpretation of Lim et al. 2018) only found in anomalous GCs, even though NGC~3201 is mono-metallic and presents a chromosome map and SGB of typical GCs. We present here our discovery on the multiple CN-CH sequences of NGC~3201.



\section{The CN-CH anti-correlation sequences}

\begin{figure}[htb]
   \centering
   \includegraphics[width=\columnwidth]{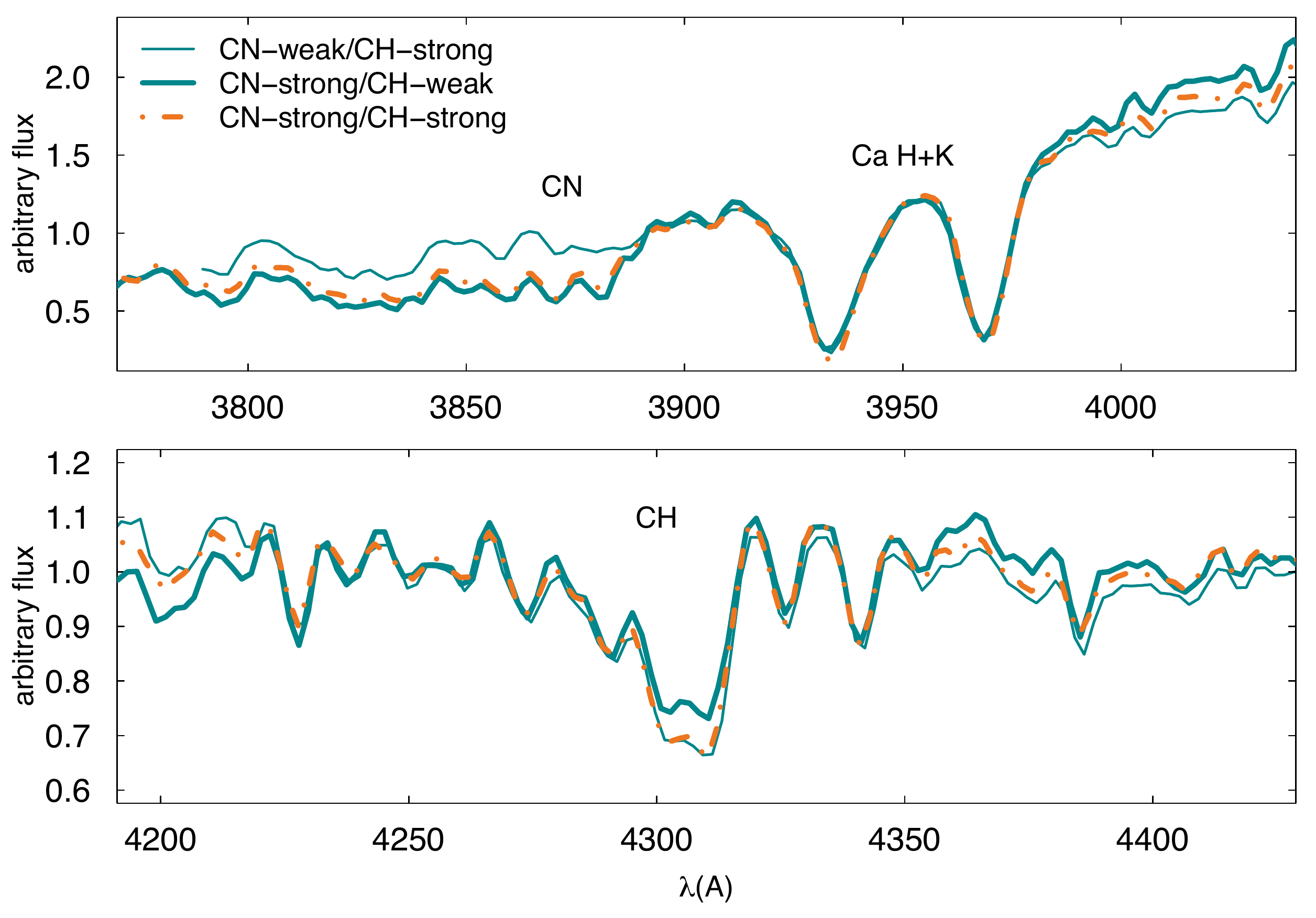}
   \caption{Spectra of three red giant stars analysed in this work, zoomed in around the CN and CH bands. The typical anti-correlated indices can be identified by the two spectra in turquoise solid lines. The orange dot-dashed line reveals a clear case of a star outside this anti-correlation.}
    \label{fig1}%
    \end{figure}

\begin{figure}[htb]
   \centering
   \includegraphics[width=0.8\columnwidth]{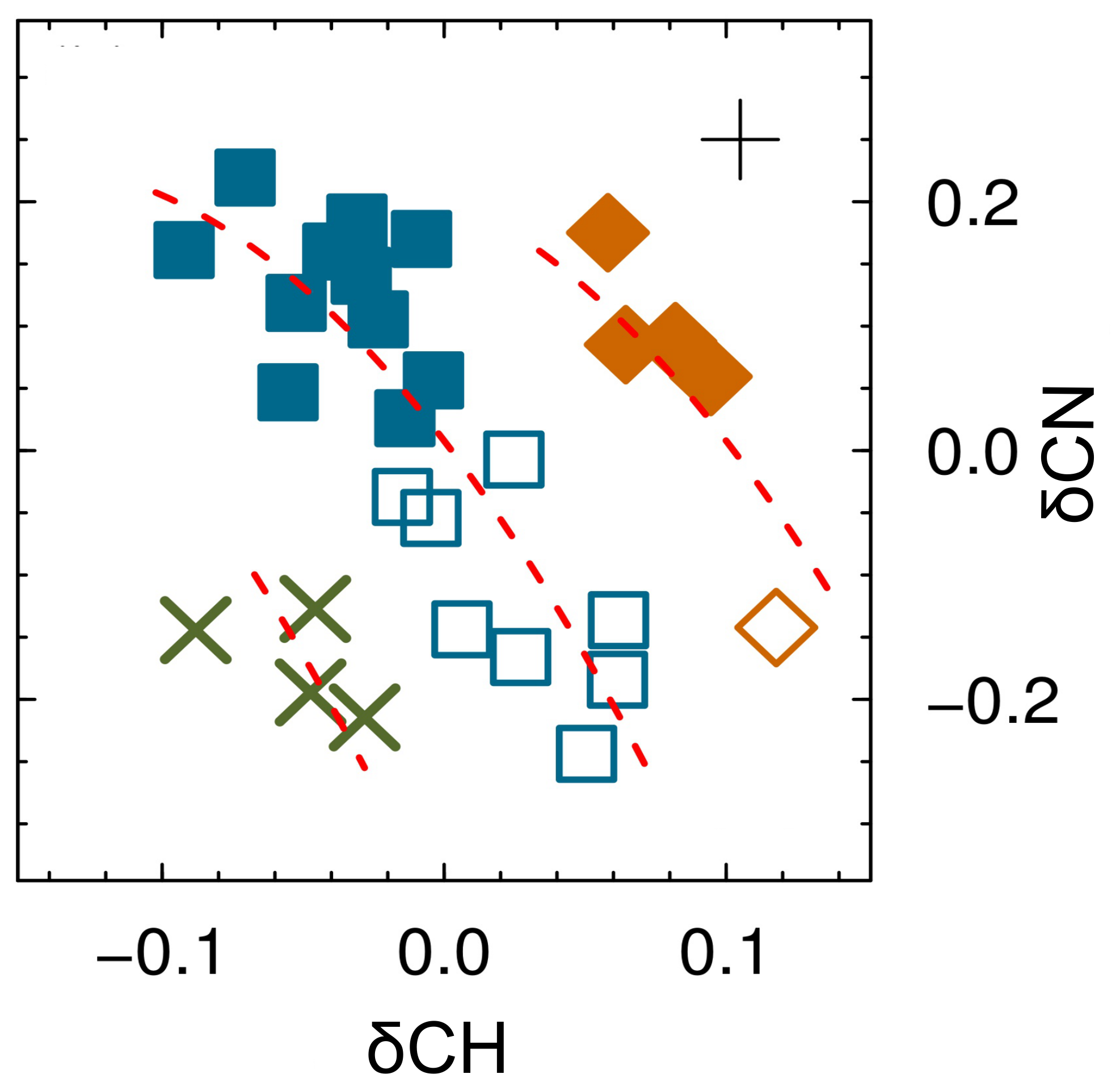}
   \caption{CN and CH indices corrected from stellar surface gravity effects for 28 RGB member stars of NGC~3201. Green crosses, blue squares, and orange diamonds represent the three sequences we found, separated by 7$\sigma$ as indicated by the error bars at the top right corner. Empty and full symbols are first and second generation stars.}
    \label{fig2}%
    \end{figure}

We observed NGC~3201 selected red giant branch (RGB) stars using the multi-object spectroscopic (MOS) mode of EFOSC2 at NTT, ESO with exposure time of one hour for each of the three pointings. The grism \#07 was used, meaning a wavelength coverage of 3270 to 5240Å, slit width 1.34", and length 8.6", resulting into a spectral resolution of $\Delta\lambda = 7.4 {\rm\AA}$, or R $\approx$ 500. From the 48 observed spectra we ended up with 28 high-quality spectra (S/N$_{CN} >$ 25). After deriving radial velocities using the Ca H,K lines as reference for cross-correlation and correcting the spectra to the rest frame, we normalized the spectra locally (see Fig. \ref{fig1}), using the local pseudo-continua defined by Pickles (1985) and Harbeck et al. (2003). We then applied the CN and CH index definition by  Harbeck et al. (2003) to measure the indices. These values are sensitive to stellar surface gravity that we corrected using magnitudes. The resulting corrected indices are shown in Fig. \ref{fig2}.

Globular clusters present a C-N anti-correlation with first generation stars being C-rich and N-poor and second generation stars C-poor and N-rich. The absolute position of this anti-correlation sequence varies with metallicity, i.e., metal-rich GCs have a sequence more enriched in carbon than metal-poor GCs (e.g. Briley et al. 2004). Anomalous clusters such as M~22 present multiple C-N anti-correlation sequences, consistent with different metallicities for the multiple groups of stars (e.g. Marino et al. 2011). NGC~3201 does not show star-to-star Fe-spread (e.g. Simmerer et al. 2013, Mu\~noz et al. 2013, Mucciarelli et al. 2015) and still we found three sequences of CN-CH anti-correlation. This discovery represents a puzzle for stellar astrophysics and globular cluster formation studies.

Marino et al. (2015) defined the photometric index $c_{BVI} = (B-V) - (V-I)$ that increases with increasing [Ba/Fe] for the NGC~5286 cluster stars, studied in that work. Even though broad band photometry should not be very sensitive to s-process elements, this relation worked well for that cluster. We used available photometry for NGC~3201 already corrected from differential reddening by Kravtsov et al. (2009) to calculate $c_{BVI}$ for all analysed stars. There is no strong correlation as in the case of NGC~5826, but there is some indication of a trend consistent with the multiple CN-CH sequences. Therefore, the multiple sequences that we found for NGC~3201 do not seem to correlate with metallicity, but at least they seem to correlate with s-process element abundances. High-resolution spectroscopic follow-up is required in order to fully characterize and understand the nature of NGC~3201 and its stars.

Photometry in ultraviolet bands was used to construct the chromosome maps of $\delta c_{F275W,I}$ versus $\delta c_{F275W,F336W,F438W}$ by Milone et al. (2017). Typical clusters shows the first and second generation stars in well-defined regions of this plot. Anomalous clusters present multiple sequences of first and second generation stars, such as NGC~1851. NGC~3201 resembles a typical cluster, and in fact the stars from Fig. \ref{fig2} appear mixed in this plot.

Our discovery raised many questions about NGC~3201. In fact, one year after Dias et al. (2018), Marino et al. (2019) raised similar questions about NGC~3201 after analysing its chromosome map in more detail.


\section{Summary}

High-resolution spectroscopic analysis of our sample stars is required to confirm or discard the evidence we discovered that the CN-CH anti-correlation groups seem to have different abundances of C+N+O and [s-process/Fe]. Three scenarios are postulated here: (i) if the sequence pec-S1-S2 has increasing C+N+O and s-process element abundances, NGC~3201 would be the first anomalous GC outside of the mass-size relation; (ii) if the abundances are almost constant, NGC~3201 would be the first non-anomalous GC with multiple CN–CH anti-correlation groups; or (iii) it would be the first anomalous GC without variations in C+N+O and s-process element abundances. In all cases, the definition of anomalous clusters and the scenario in which they have an extragalactic origin must be revised.


\end{document}